\documentclass[12pt]{iopart}
\usepackage{braket}
\usepackage{graphicx}

\begin{document}

\title{A Bell telegraph}

\author{\large Daniel Badagnani$^{1,2}$}
\address{$^1$Departamento de F\'\i sica,
Facultad de Ciencias Exactas, Universidad Nacional de La Plata,
C.C. 67, 1900 La Plata, Argentina.}
\address{$^2$UIDET-CETAD, Facultad de Ingeniería, Universidad Nacional de La Plata}

\ead{daniel@fisica.unlp.edu.ar}



\begin{abstract}

We show a device with which, apparently, information (in the form
of a "slash-dot" code) is instantly transmitted via Bell state collapse,
over arbitrary distance. We show then why the device does not actually work, and discuss about
the ``no Bell telephone'' statement. We discuss also about the troublesome relationship 
between the Copenhagen interpretation
and special relativity. In particular, we advocate the interpretation that the ``state space''
cannot be a description of the state of the object studied alone, but that the observer
is inextricably involved even if no measurement is taking place.

\end{abstract}

\section{Introduction}

``There are no Bell telephones'', says an old joke in the community of foundations
of quantum mechamics, meaning that one cannot (should not) be able to use Bell's
nonlocality (collapse of entangled quantum states at a distance due to local measurements, of
the kind of the EPR argument \cite{EPR,Bell})
for transmitting information. There is no general proof of this statement, and that's fortunate
since Bell himself claimed once that all an impossibility theorem
proves is lack of imagination. However, if quantum mechanics admitted such information
transmission to be performed, it would enter in direct conflict with very firmly established
consequences of special relativity. In such a case one of these theories (or both) would be in serious
trouble, and the main suspect would be the Copenhagen interpretation of quantum mechanics.

What we present here is an apparent ``Bell telegraph'' and the reasons why it does not work. It is
a replacement of a ``youth sin'' version from 2002, in which I posed it in a ``paradox format'',
expecting to spark discusions and reactions (I did, indeed).
But, to my surprise, the manuscript attracted also some attention from the media, 
where it was regarded not as a theoretical
puzzle to be solved but as an actual ``invention'', a proposal to actually communicate instantly at
arbitrary distances \footnote{see for instance http://www.freerepublic.com/focus/news/735915/posts, 
last accessed 10/4/2017}. I hope this version will frame the discussion correctly
as an effort in thinking about the interpretations of quantum mechanics.

This manuscript is organized as follows: the telegraph is first presented exactly as in the 2002 version.
In the next section (also exactly that of the 2002 version) the conflict with special
relativity is discussed. Then, in section (4) (which is new) an explanation of 
why the telegraph does not work is shown (you might challenge yourself trying to understand it before reading
that section). We also briefly discuss to what extent the `no Bell telephone''
conjecture can be considered a feature of quantum mechanics under the Copenhagen interpretation. 
Finally, we discuss the description of collapse in a relativistic realm and pose some questions.

\section{The device}

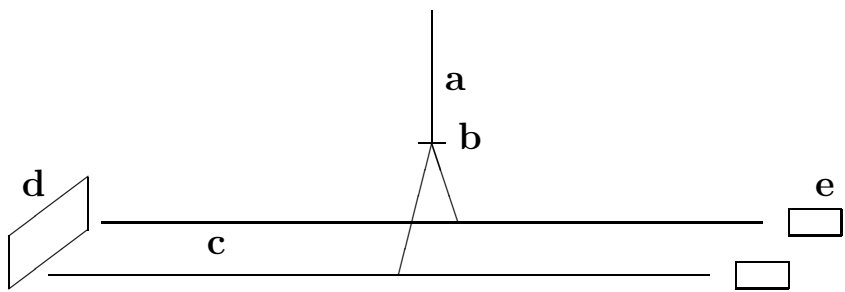
\begin{figure}
\begin{picture}(330,100)(0,0)

\put(30,0){\line(1,0){250}}
\put(90,8){\large \bf c}

\put(290,-5){\line(0,1){10}}
\put(290,5){\line(1,0){20}}
\put(290,-5){\line(1,0){20}}
\put(310,-5){\line(0,1){10}}

\put(175,50){\line(0,1){50}}
\put(180,70){\large \bf a}

\put(170,50){\line(1,0){10}}
\put(185,48){\large \bf b}

\put(162.5,0){\line(1,4){12.5}}
\put(185,20){\line(-1,3){10}}

\put(50,20){\line(1,0){250}}

\put(310,15){\line(0,1){10}}
\put(310,25){\line(1,0){20}}
\put(330,15){\line(0,1){10}}
\put(310,15){\line(1,0){20}}
\put(320,30){\large \bf e}

\put(15,-5){\line(4,3){30}}
\put(15,15){\line(4,3){30}}
\put(15,-5){\line(0,1){20}}
\put(45,17){\line(0,1){20}}
\put(20,30){\large \bf d}

\end{picture}

\begin{center}
\caption{Scheme of the telegraph. A beam of excited atoms (a) splits
at a potential barrer (b). Then the atom decays at any of two pipes (c)
emitting two correlated photons. At one end (d) they are set to interfere
with themselves at a screen. At the other end (e), two detectors are
placed, which can be turned on or off}
\end{center}
\end{figure}

A sketch of the device is shown in the figure above. It is inspired
in Feynman's discussion on quantum interference \cite{Feynman}. A beam of
excited atoms (which will emit a two photon Bell state) is divided in two by
a potential barrier in such a way that a given atom has the same probability
of following each path, which is itself a Bell state. At the decay, the
resulting photons are collected by some kind of pipes (for instance, two
optic fibers), so every pair of photons is in any of those, forming a kind
of doubly entangled state. If no one collapses this system, the photons at
one end of the pipes can be set to interfere at a screen. But one could
'see' where each pair is just looking at the second photon at the other end,
destroying the interference pattern. So, turning on and off detectors at one
end sets the other end at two observably different states (random arrival
and interference respectively), which can obviously be used as a telegraph.

Still we have to show that this can indeed be used to send information in a
time as short as necessary to be considered instantaneous. We need several
photon pairs (say M) to conclude that we are before an interference or not,
with a given confidence interval. One solution is to use many uncorrelated
pairs simultaneously, and invoke superposition principle. But in order not
to obscure the conclusions by adding hypothesis, we will give an alternative
proof using single pairs. Suppose that for a given telegraph we need a time
separation T for producing each pair being sure that we deal with a single
pair at a given instant. So we need a time MT for sending a "byte". Now take
N such telegraphs, in which the first pair in each is produced in an instant
at random between 0 and T. With this ensemble we build an improved
telegraph, turning on and off all detectors simultaneously for sending
signals. Now the time needed reduces to MT/N, and can be made as little as
we wish. Thus we can send information instantly (at the large N limit) using
(lots of) single pair states.

\section{Troubles}

Let's suppose for the moment that we haven't make any mistakes; if so,
quantum mechanics imply we can send information instantly. What do we mean
by "instantly" in a relativistic context? Or, put differently, instantly in
what reference frame?  This is a difficulty with the very
concept of quantum collapse. Even if it turns that no information could
be transmitted via Bell states, this is still a conceptual puzzle.
Whatever the answer is, it fits into either of the following two posibilities:

\begin{itemize}
\item  {{\bf Deny relativity:} There is a privileged frame F in which
quantum collapse occurs instantly. This possibility is not inconsistent but
deeply nasty.}

\item  {{\bf Preserve relativity:} The frame F in which the collapse is
instantaneous depends on the state (for instance, it could be the one in
which the center of mass is at rest). This possibility seems more
acceptable, but a closer look shows it leads to inconsistencies, which can
be seen as paradoxes. Consider the setting shown in the figure below: two
identical telegraphs A and B with F(A) and F(B) moving in opposite
directions results in sending a signal to the past. As F(A) moves to the
left, A reception happens before A emmision in this frame. The message is
instantly retransmitted trough B (B emmision) and again, as F(B) moves to
the right, B reception happens before B emission, and at the same space location
than A emmision. In the setting, the
automaton located there transmit $m_1$ if and only if it transmit $m_2$ and
vice versa, which is absolutely inconsistent.}

\end{itemize}

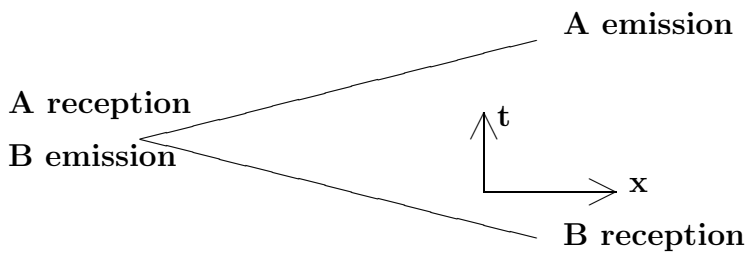
\begin{figure}
\begin{picture}(300,100)(0,0)

\put(50,50){\line(4,1){150}}
\put(0,60){\bf A reception}
\put(0,40){\bf B emission}
\put(50,50){\line(4,-1){150}}
\put(210,90){\bf A emission}
\put(210,10){\bf B reception}

\put(180,30){\line(1,0){50}}
\put(230,30){\line(-2,1){10}}
\put(230,30){\line(-2,-1){10}}
\put(235,30){\bf x}
\put(180,30){\line(0,1){30}}
\put(180,60){\line(1,-2){5}}
\put(180,60){\line(-1,-2){5}}
\put(185,55){\bf t}

\end{picture}

\caption{Setting for the paradox: two identical and opposely oriented
tellegraphs A and B have F(A) and F(B) respectively as frames where
communication is instantaneous. F(A) moves toward the left at some
speed v, while F(B) is set to move towards the right at speed v, and we
produce the A emission in a moment such that A reception and B emission
coincide. The message received from A is retransmitted through B. An
automaton able to send either message $m_1$ or $m_2$ stay still at the
position of A emission, which is the same than B reception, and we set it to
send $m_1$ throug A if it reads $m_2$ from B and vice versa.}
\end{figure}

Of course, there remains the possibility that we have made some mistake and
quantum mechanics really forbids such a device to work. Hopefully this work
will call someone's attention in order to find it out.

\section{The resolution and the ``astuteness'' of the Copenhagen interpretation}

When calculations are carried carefully, the result is that there is never interference in the screen,
irrespective of what is done in the detectors. Call the ``pipes'' $1$ and $2$, call $L$ the side of the screen 
and $R$ the side of the detectors. Observe that prior to any measurement, the quantum
state of the system is
\begin{equation}
\Ket{\Psi_0} = \frac{1}{\sqrt{2}} \left( \Ket{1L}\otimes\Ket{1R} + \Ket{2L}\otimes\Ket{2R} \right)
\label{priorstate}
\end{equation}
where $\Ket{iL}$, $\Ket{jR}$ are supposed to be normalized, localized, time dependent and mutually orthogonal
state vectors. If we call $\Ket{x}$ the eigenstates of position on the screen, the amplitude on it will read
\begin{equation}
\Braket{x | \Psi_0} = \frac{1}{\sqrt{2}} \left( \Braket{x | 1L}\Ket{1R} + \Braket{x | 2L}\Ket{2R} \right)
\label{StatScreen}
\end{equation}
which is still a state vector. The probability density of finding a photon in a position on the screen is thus
its modulus:
\begin{eqnarray}
p(x) &=& \frac{1}{2} \left(
          |\Braket{x | 1L}|^2 \Braket{1R | 1R} + |\Braket{x | 2L}|^2 \Braket{2R | 2R}
                   \right) \label{probacruda} \\
     &=& \frac{1}{2} \left(|\Braket{x | 1L}|^2+|\Braket{x | 2L}|^2\right)
\label{probastuta}
\end{eqnarray}
where we used the orthonormality of the vectors $\Ket{1R}$ and $\Ket{2R}$. Thus, even if we do not collapse
the wavefunctions with the detectors, the pattern is just a statistical average showing no interference.
Of course, if the detectors on the right operate, the state at the moment of the screening will not be
(\ref{StatScreen}); assume the detectors measure the observable O, and that there is a complete ortonormal basis
for the observables $\Ket{O^{(1)}_k}$ and $\Ket{O^{(2)}_{k'}}$. If the outcome is for instance the eigenvalue $O^{(1)}_l$,
then the state will be 
$$
\frac{1}{\sqrt{2}}\Braket{x | 1L} \Braket{O^{(1)}_l | 1R} \Ket{O^{(1)}_l},
$$
which will
hapen with (the conditional) probability 
$$
P\left(x | O^{(1)}_l \right) = \frac{1}{2} \|\Braket{x | 1L}|^2 \|\Braket{O^{(1)}_l | 1R}\|^2.
$$
But since at the screen there is not
information about what happened at the detectors, the probability of finding the photon at $x$ will be the sum on all
outcomes on the detector, and using the completeness of the base and its orthonormality the probability (\ref{probastuta})
is recovered.

It should be noticed that the ``no Bell telegraph'' feature is far from being built-in the formalism, which is indeed
highly non-local, but at the same time very astute. It comes as somewhat of a miracle that the entanglement, which provides
the possibility of influencing at a distance, provides also an orthogonality preventing the system of making observability
in one end dependent on what is done in the other. To appreciate the extent of this astucy, let us consider the following
``upgraded telegraph'': instead of trying to destroy interference with the detectors, we try to recover the interference by
re-joining the path of the $R$-side photon with a system of mirrors, with the hope that ending with the physical separation
we will make the $R$-states non-orthogonal. This does not work, provided the evolution operator is unitary, thus preserving
the inner products in the state space. This, in turn, is a consequence of the hermiticity of the Hamiltonian, but an actual
system of mirrors will be slighty absorptive, and thus non-Hermitian. ¿Can we be absolutely positive that any conceivable
system that recombinates the states $\Ket{1R}$ and $\Ket{2R}$ will preserve their orthogonality?

\section{So, is collapse compatible with special relativity?}

Let us suppose that we take for granted that there is no Bell telegraph. We will not find then any observational inconsistency
between quantum mechanics and special relativity. But the description of the states will change radically in different reference 
systems. There is no question that, in a given reference frame, the collapse is to be described as instantaneous. But then
what measurement collapses the state will depend on the observer, and it seems impossible to avoid this ugly non-covariance
at the level of the description of quantum collapses. We might wonder if we aren't mixing theoretical frames by placing
entanglement and collapse in a relativistic limit, but we can see easily that such a caveat is baseless: relativistic
quantum field theory is a standard quantum theory whose state space can be construed as a system of an unlimited number of 
identical particles, and so entangled states are part of the state space of the relativistic field. Measurement in the
context of relativistic fields has been pursued in the heydays of quantum theory, but to our knowledge it has not
been developed in order to incorporate the developements involving entanglement.

\begin{figure}\begin{center}
  \includegraphics[width=12cm]{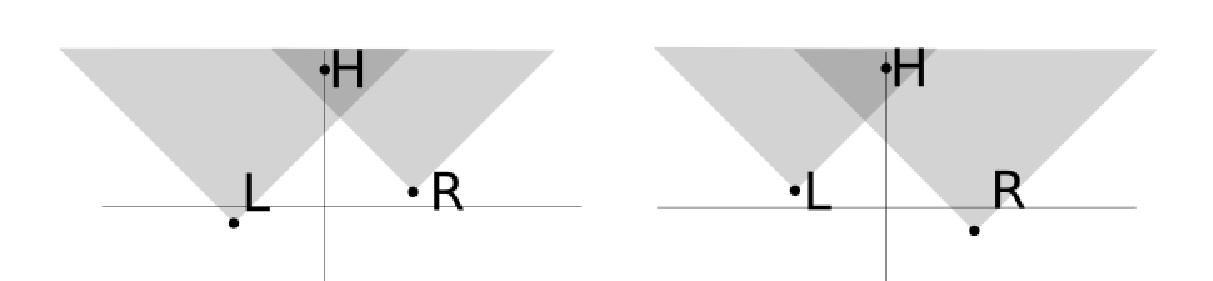}
  \caption{Measurements in the laboratories $L$ and $R$ in different reference frames, with the respective future light cones.}
  \label{historian}
  \end{center}
\end{figure}

Let us consider two remote detectors measuring a system of entangled particles in two laboratories $L$ and $R$ separated
by a space-like four-vector, as shown in figure (\ref{historian}). A historian $H$ is placed in a common region of the future
light cones of $R$ and $L$.

An entangled state is prepared in the state $\Ket{L}\otimes\Ket{R}$. At $L$ a detector will measure one from the
complete set of states $\Ket{O^{(L)}}_k$, while at $R$ another detector will measure one from $\Ket{O^{(R)}}_{k'}$.
In terms of these states, the entangled state can be cast in the form
$$
\left( \Sigma_k \Braket{O^{(L)}_k | L} \Ket{O^{(L)}_k} \right) 
                           \otimes 
\left( \Sigma_{k'} \Braket{O^{(R)}_{k'} | R} \Ket{O^{(R)}_{k'}}  \right)
$$
Let us suppose the historian determines that at $L$ the value $O^{(L)}_A$ has been obtained, while on the left
the outcome is $O^{(R)}_B$. He will thus conclude that the state after both laboratories performed their measurements
is $\Ket{O^{(L)}_A} \otimes \Ket{O^{(R)}_B}$. What was the state at time zero in the figure?

In the figure of the left the movement of the observer is such the measurement at $L$ happen at $t<0$, while measurement 
at $R$ is at $t>0$. Thus, at time zero the state is:
$$
\Ket{O^{(L)}_A} \otimes \left( \Sigma_{k'} \Braket{O^{(R)}_{k'} | R} \Ket{O^{(R)}_{k'}}  \right).
$$
On the other hand, for an observer which moves such that the order is the one shown on the right, the state at time zero is
$$
\left( \Sigma_k \Braket{O^{(L)}_k | L} \Ket{O^{(L)}_k} \right) \otimes \Ket{O^{(R)}_B}.
$$
So, can these be actually the states of the system? How can we understand that such state will depend on the movement of the
observer? Even if this highly non-covariant description shows no observable non-covariance, it seems difficult to state
that the vectors shown correspond to the states of the system under study. It might be more proper to think that it is
impossible to separate system from observer, even while no measurement is being performed.

\section{Concluding remarks: why should we care about all this?}

We could separate the caveats about collapsing entangled states in ``hard'' (observable) possible covariance
 problems and ``soft''
(interpretational) worries. If one takes a positivist point of view, only the hard problems should bother us. As we have seen,
quantum theory plus Copenhagen interpretation seems astute and, in spite of permitting conspicuous non-local and 
non-covariant descriptions at the unobservable level of the states, preparing a system that violates covariance at the level
of the observables is seemingly impossible. We could try to prove stringent ``no-go'' theorems, but history warns us that
it is impossible to have control over hidden lemmas: conditions assumed implicitly in the proof of such theorems.
It is in this sense, I think, that the celebrated quote ``all an impossibility theorem proves is lack of imagination'' should
be interpreted. If the theory actually had such ``hard'' non-covariance, it should be interpreted, I think, as a formal
difficulty of quantum mechanics, since relativistic causality is so firmly established, and not as a prediction of the
technical possibility of superluminal communication. It should be said that the possibility of such ``hard'' noncovariance
seems extremely remote.

However, the ``metaphysical'' soft non-covariance should not be underestimated. Recall that the Viena circle failed
in rooting all meaning on ``empiric facts'' \cite{VC}.
It is important to bear in mind
that quantum collapse remains the most difficult problem in the interpretation of quantum mechanics, and understanding it
is unavoidable if the potential practical applications of entanglement (quantum computing, teletransportation, etc) are
to work at a scale large enough to be more than expensive toys. In this sense, wondering ``what is the state of a system''
when the sequence order of collapse is frame-dependent is indeed an intriguing problem. I feel that holding the belief that
the state is strictly a property of the system is untenable. Our examples of remote collapse seemingly show that the state vector
is a description of the state of the system-observer complex, becoming impossible to separate subject and object even while
measurements are not being performed.

\section{Aknowledgements}

I am gratefull to Alejandro Daleo and Monica Mance\~nido for very useful comments and discussions.
I am also grateful to the many that wrote after the 2002 version was released.
Regrettably, I did not preserve those e-mails, not even the original e-mail box.
I deeply regret to have waited fifteen years to make this replacement. The delay
has to do with many personal and national events here in my country.

\end{document}